\documentclass[twocolumn,final,showpacs,
               showkeys,
               aps,superbib,prb,
               letterpaper,balancelastpage,
               lengthcheck,byrevtex,bibnotes,
               footinbib,
               ]{revtex4}
\usepackage{graphics}
\usepackage{graphicx,natbib,amssymb}
\usepackage[thinspace,squaren]{SIunits}
\def\ie{{\em i.e.}}

\def\full{\protect\mbox{------}}
\def\kesik{\protect\mbox{--\, --\, --}}
\def\chain{\protect\mbox{-- $\cdot$ --}}
\def\drt{{distribution of relaxation times}}
\def\hnef{{empirical formula}}

\def\full{\protect\mbox{------}}
\def\kesik{\protect\mbox{-\, -\, -\, -}}

\begin{document}

\title{Origin of temperature dependent conductivity of $\alpha$-polyvinylidene fluoride}
\author{\firstname Enis \surname Tuncer}
\email{enis.tuncer@physics.org}
\author{\firstname Michael \surname Wegener}
\author{\firstname Peter \surname Fr{\"u}bing}
\author{\firstname Reimund \surname Gerhard-Multhaupt}
\affiliation{Applied Condensed-Matter Physics, Department of Physics, University of Potsdam, D-14469 Potsdam Germany}
\date{\today}

\begin{abstract}
The conductivity of $\alpha$-polyvinylidene fluoride ($\alpha$-PVDF) is obtained from dielectric measurements performed in the frequency domain at several temperatures. At temperatures above the glass-transition, the conductivity can be interpreted as an ionic conductivity, which confirms earlier results reported in the literature. Our investigation shows that the observed ionic conductivity is closely related to the amorphous phase of the polymer.
\end{abstract}
\maketitle
One advantages of highly insulating polymers is their low conductivity. 
It is therefore of interest to understand the conduction mechanisms in these materials. In this short letter, we present the conductivity data of $\alpha$-polyvinylidene fluoride ($\alpha$-PVDF) as a function of temperature. $\alpha$-PVDF is a semi-crystalline polymer used in various technological applications~\cite{TashiroFerro}. 
In order to better understand and design systems with PVDF for special applications, the influence of the temperature on various material properties must be known. 
\citet{Eberle1996} and the \citet{Ieda1984} group performed pressure-dependent conductivity measurements, and concluded that the conductivity in $\alpha$-PVDF is ionic. They mention that the free-volume change with pressure or at the glass transition strongly influences the ionic conductivity. As a result, the ionic conductivity decreases with increasing pressure because of decreasing free volume and mobility of polymer segments. Here, we employ the dielectric-spectroscopy technique to investigate the conduction behavior of $\alpha$-PVDF.

The conduction in polymers is in general expressed as a function of temperature in the form of a stretched exponential~\cite{Chauvet1992,Conner1998},
\begin{eqnarray}
  \label{eq:2}
  \sigma=\sigma_0 \exp(-T_1\, T^{-\alpha}),
\end{eqnarray}
where $\alpha$ indicates which type of conduction is dominant in the system; $\alpha=1$ means that the conduction is activated over an energy gap, a barrier, or a mobility edge [for $\alpha=1$, Eq.~(\ref{eq:2}) becomes identical to Arrhenius expression with $T_1=E/k$ (activation temperature), and $E$ activation energy; $k$ is the Boltzmann constant ($=86.1321\ \micro{e}\volt\reciprocal\kelvin$)], while $\alpha=1/4$ is considered to point towards variable-range hopping of localized carriers, and $\alpha=1/2$ to Coulomb interaction between the carriers. In addition, it is suggested~\cite{Sheng1978} that $\sigma$ in disordered systems can be related to tunneling through a potential barrier of varying height due to local temperature fluctuations. In this case, the conductivity is expressed as in the form of a Vogel-Fulcher-Tammann (VFT)~\cite{Vogel} equation with a critical temperature $T_0$,
\begin{eqnarray}
  \label{eq:3}
  \sigma=\sigma_0 \exp[-T_1(T-T_0)^{-1}],
\end{eqnarray}
where $T_1$ and $T_0$ are related to physical dimensions of microscopic system 
and 
the barrier height a charge carrier has to overcome (for  details see Refs.~\onlinecite{Conner1998} and \onlinecite{Sheng1978}). 
\begin{figure}[t]
  \begin{center}
    \includegraphics[width=\linewidth]{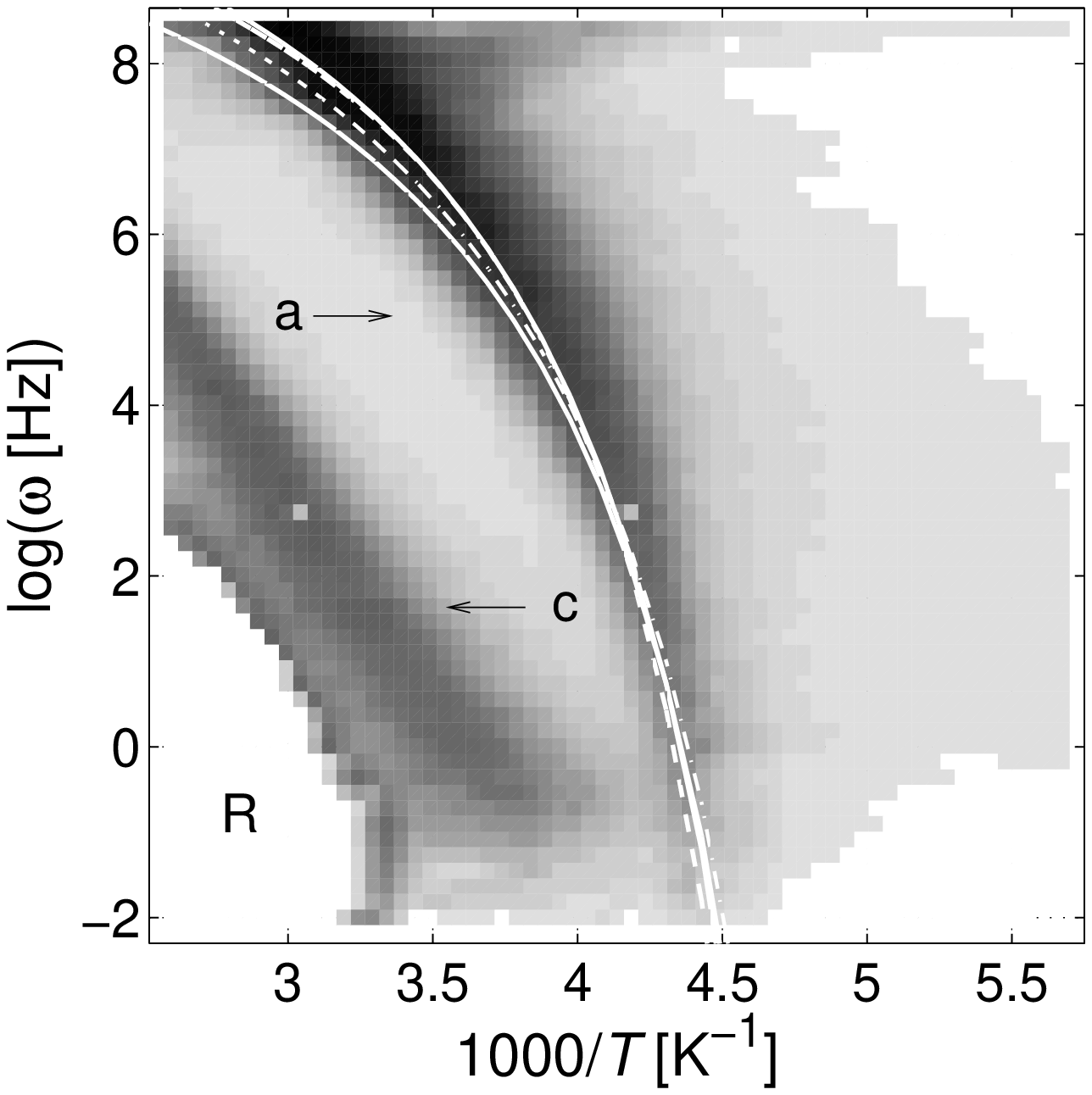}
      \caption{Relaxation map for $\alpha$-PVDF with amorphous `{\sf a}' and crystalline `{\sf c}' phases. Two solid (\full) and dashed (\kesik) lines represent VFT fits obtained from the \drt\ and the Havriliak-Negami \hnef, respectively. The results from Ref. \onlinecite{Grimau1999} are plotted as chain lines (\chain). The gray scale illustrates the probability density of relaxations ${\sf g}(\tau)$, dark areas are most probable regions for the particular relaxation. The omitted region, labeled `{\sf R}', contains information related to melting of the sample. \label{fig:relaxmap}}
  \end{center}
\end{figure} 

In ionic glass-former conductors, $\sigma$ is best described by an equation similar to Eq.~(\ref{eq:3})~\cite{Wang1997}
\begin{eqnarray}
  \label{eq:5}
  \sigma=\sigma_0{T}^{-1/2}\exp[-T_1(T-T_0)^{-1}],
\end{eqnarray}
which is derived from VFT expression in Eq.~(\ref{eq:3}) with multiplication of Nernst-Einstein equation, which  transform $\sigma_0$ in Eq.~(\ref{eq:3}) to $q^2nD/kT$, where $q$ is the charge of an ion, $n$ is the ion concentration and $D$ is the diffusion coefficient ($\propto T^{1/2}$). In the last expression, $T_0$ is the temperature at which the mobility of the ions disappears. For glass-forming amorphous ionic conductors critical temperature $T_0$ should be related to the glass-transition temperature $T_g$, $T_0\simeq T_g$. For amorphous polymers, the ionic conductivity is affected by the cooperative motion of polymer-chain segments~\cite{Wang1997}. 
\begin{table}[b]
\caption{Fit parameters for the `{\sf a}' relaxation. $T_g$ is determined as the temperature at with the VFT fits [Eq.~(\ref{eq:3})] yield $\log\omega=-2$. The abbreviations drt and hn in the `Note' column indicate the \drt\ and the Havriliak-Negami \hnef, respectively.\label{table}}
\begin{tabular*}{\linewidth}{@{\extracolsep{\fill}}lrrllc}
\hline
  & $\sigma_0~[\times10^{12}]$ & \multicolumn{1}{r}{$T_1$} & \multicolumn{1}{r}{$T_0$} & \multicolumn{1}{r}{$T_g$}&Note\cr
\hline
This work & 39.0 & 918 & $191$ &   223 &drt \\
This work & 6.0 & 988 &  $191$ &  224 & drt \\
This work & 12.0 & 886 &  $196$ & $225$ & hn \\
Ref.~\onlinecite{Grimau1999}& 12.0 &  $1021$ & $187.5$ &  222.1& hn \\
\hline
\end{tabular*}
\end{table} 

In addition, an expression, which was resulted from the stretched exponential relaxation-time distribution within a defect diffusion model, is proposed for the ionic conductivity~\cite{Bendler2001}
\begin{eqnarray}
  \label{eq:6}
  \sigma=\sigma_0\exp[-T_1(T-T_0)^{-3/2}],  
\end{eqnarray}
where $\sigma_0$, $T_1$ and $T_0$ are all nearly constant. As the material is cooled down, the number of defects is lowered and the material becomes more viscous, thus the conductivity decreases.
 
Here, we adopt the expressions in Eqs.~(\ref{eq:3}),~(\ref{eq:5}) and (\ref{eq:6}). All of these expressions have three adjustable parameters, namely $\sigma_0$, $T_1$ and $T_0$.  
It is observed that the expressions yield similar results which cannot be distinguished easily. We therefore discuss the physical meaning of the parameters from the curve fitting results together with the residuals ${\sf \chi_e}$,
\begin{eqnarray}
  \label{eq:7}
  {\sf \chi_e}=\sum[\log(\sigma)-\log(\sigma_e)]^{2}
\end{eqnarray}
where $\sigma_e$ are the conductivities as calculated from a \drt\ in dielectric data, and $\sigma$ are the values predicted by Eqs.~(\ref{eq:3}),~(\ref{eq:5}) and (\ref{eq:6}).

The electrical properties of the investigated $\alpha$-PVDF were measured with the  broadband dielectric spectroscopy. The crystallinity, calculated from the values of the experimental melting entalpy ($42\ \joule\reciprocal\gram-50\ \joule\reciprocal\gram$) and the equilibrium melting entalpy of a perfect crystal of PVDF\cite{Nakagawa1973} ($=104.7\ \joule\reciprocal\gram$), of the samples were between $40\%$ and $50\%$. The melting point of the samples were around $178\ \celsius$. 
In Ref. \onlinecite{Grimau1999}, the dielectric properties of $\alpha$-PVDF have been reported for the same frequency intervals, but at lower temperatures ($T<285\ \kelvin$), and the conductivity contributions to the dielectric spectra were disregarded. In our experiments, the frequency and temperature intervals were between $10^{-1}-10^{6}\ \hertz$ and $173-430\ \kelvin$, respectively. The dielectric data were analyzed with two different approaches, fitting (i) a sum of Havriliak-Negami empirical functions~\cite{HN} and (ii) a distribution of relaxation times~\cite{Tuncer2004a}.

\begin{figure}[t]
  \begin{center}
    \includegraphics[width=\linewidth]{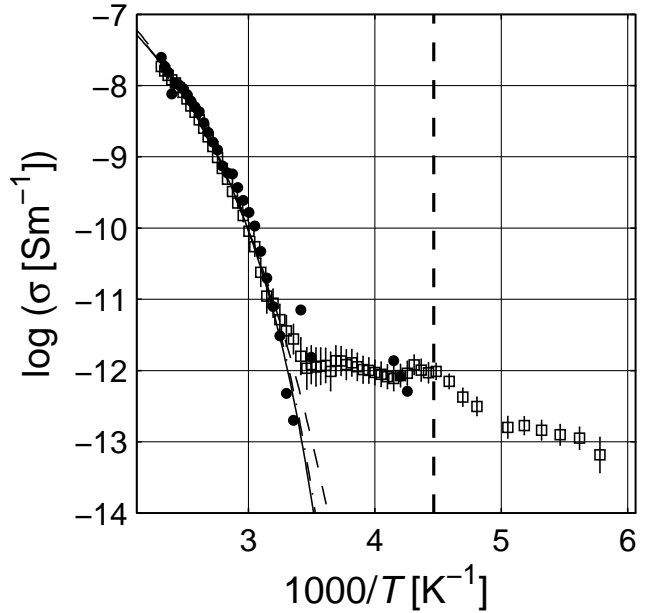}
      \caption{Conductivity of $\alpha$-PVDF as a function of temperature. The open squares with error bars ($\Box$) and the full circles ($\bullet$) represent the values obtained from the \drt\ and from the Havriliak-Negami \hnef, respectively. Solid (\full), dashed (\kesik) and chained (\chain) lines are curve fits to Eqs.~(\ref{eq:3}), (\ref{eq:5}), and (\ref{eq:6}) respectively. The thick vertical dashed line indicates the glass-transition temperature $T_g$. \label{fig:sigma}}
  \end{center}
\end{figure} 
\begin{figure}[t]
  \begin{center}
    \includegraphics[width=\linewidth]{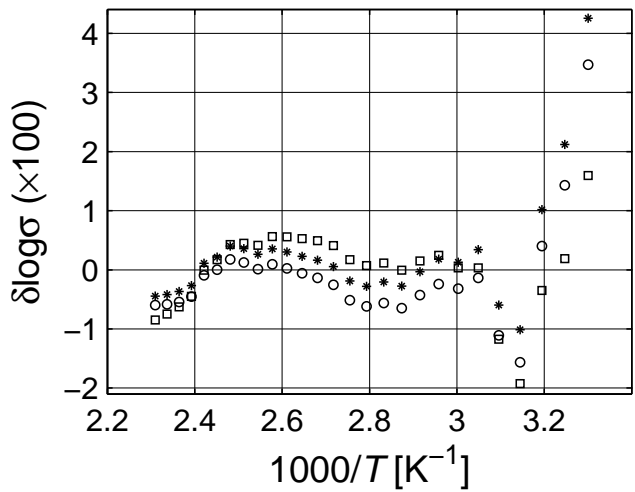}
      \caption{
Relative error for the three curve-fitting procedures. The symbols ($*$), ($\circ$) and ($\Box$) represent the errors calculated for Eqs.~(\ref{eq:3}), (\ref{eq:5}) and (\ref{eq:6}), respectively.\label{fig:sigmares}}
  \end{center}
\end{figure} 

\begin{table}[b]
\caption{Fit parameters for the conductivity. \label{table2}}
\begin{tabular*}{\linewidth}{@{\extracolsep{\fill}}lrrrrc}
\hline
  & $\sigma_0~[\times10^{6}]$ & $T_1$ & $T_0$ & ${\sf \chi_e}$& Note\\
\hline
This work & 7.45 &$1240$ & $224$ & 0.335 &Eq.~(\ref{eq:3})\\
This work & 20.0 &$1300$ & $221$ & 0.122 &Eq.~(\ref{eq:5})\\
This work & 6.04 &$25700$ & $158$ & 0.253 &Eq.~(\ref{eq:6})\\
\hline
\end{tabular*}
\end{table}

In Fig.~\ref{fig:relaxmap}, the relaxation map  of $\alpha$-PVDF extracted from the dielectric data by means of a \drt\ is shown. In the figure, the gray scale illustrates the probability density ${\sf g}(\tau)$ of the respective relaxations. Two relaxations  labeled `{\sf a}' and `{\sf c}' are significant, and correspond to the amorphous and crystalline regions of the polymer, respectively~\cite{Grimau1999}. The region marked `{\sf R}' contains information related to the melting of $\alpha$-PVDF with higher probability density than the other two. These relaxations with high magnitudes are disregarded in this study. In addition, the relaxation of crystalline phase `{\sf c}' is not discussed here because there is no relationship to the conductivity data. The lines in Fig.~\ref{fig:relaxmap} represent VFT fits [Eq.~(\ref{eq:3})] to the relaxation of the amorphous phase. The solid (\full) and the dashed (\kesik) lines are calculated with the \drt\ and with the Havriliak-Negami \hnef, respectively. The chain line (\chain) is taken from Ref. \onlinecite{Grimau1999} for the `{\sf a}' relaxation. In Table~\ref{table} fit parameters for  the relaxation of amorphous phase are listed. The \drt\ analysis yields asymmetric probability distributions. Therefore two different curves (presented with solid lines in Fig.~\ref{fig:relaxmap}) are used for each of these probability densities. The \drt\ and the Havriliak-Negami \hnef\ both result in similar values as the ones reported by \citet{Grimau1999}. From the VFT curves, the glass-transition temperature $T_g$ of $\alpha$-PVDF is calculated as $\sim224\ \kelvin$. $T_g$ is defined as the temperature at which the VFT fits yield $\log\omega=-2$.

In Fig.~\ref{fig:sigma}, the conductivity of $\alpha$-PVDF is shown as a function of reciprocal temperature. The Havriliak-Negami \hnef\ does not allow us to resolve conductivity values at low temperatures. Therefore, we use the data from the \drt\ instead. The conductivity of the material is increasing with a moderate slope at temperatures below $T_g$, shown with the vertical dashed line (\kesik) in Fig.~\ref{fig:sigma}. At temperatures lower than $290\ \kelvin$, $\sigma$ is weakly temperature-dependent, because the movement of charge carriers is not assisted by the cooperative motion of polymer chains. At temperatures over $290\ \kelvin$, the cooperative motion determines the movement of the charge carriers. 
The analysis of the conductivity data at high temperatures with the three expressions of Eqs.~(\ref{eq:3}),~(\ref{eq:5}) and (\ref{eq:6}) produces perfect agreement with the data. The curve-fitting parameters are presented in Table~\ref{table2}, and the relative error $(\log\sigma/\log\sigma_e -1$) is illustrated in Fig.~\ref{fig:sigmares}. 
The critical temperatures $T_0$ obtained for Eq.~(\ref{eq:5}) and (\ref{eq:6}) correspond to the glass-transition temperature of the polymer as calculated independently from the dielectric relaxation data. Although the residual of Eq.~(\ref{eq:6}) is lower than that of the VFT equation of Eq.~(\ref{eq:3}), the critical temperature is not related to $T_g$. Therefore, of the three avaliable expressions, Eq.~(\ref{eq:5}) best describes the conductivity of $\alpha$-PVDF. 
Last but not least, the conductivity values at room temperature and at $350\ \kelvin$ are close to the values reported by \citet{Fedosov2002} ($\sigma(T@300\ \kelvin)\approx10^{-11}\ \siemens\reciprocal\meter$) and by \citet{Ieda1984} ($\sigma(T@350\ \kelvin)\approx10^{-10}\ \siemens\reciprocal\meter$), respectively. The temperature dependence of the conductivity resembles that of the amorphous-phase `{\sf a}' relaxation in $\alpha$-PVDF, which has also been found for other polymeric systems~\cite{Wang1997}. 

In conclusion, our investigations by means of dielectric spectroscopy confirm the findings of \citet{Eberle1996} and \citet{Ieda1984} that the conductivity of $\alpha$-PVDF is ionic at temperatures well above $T_g$. The ions probably stem from impurities. The observed temperature dependence of the conductivity shows similarities to that of semiconductors~\cite{AshcroftandMermin}, \ie\ two separate regions (intrinsic and impurity) are observed in the conductivity as a function of temperature. A weakly temperature-dependent (impurity) conductivity is visible below $290\ \kelvin$, which is not assisted by the molecular motion. The (intrinsic) conductivity of $\alpha$-PVDF appears to be ionic at temperatures higher than $290\ \kelvin$, and is closely related to the molecular mobility above the glass transition. The ionic conductivity is thus probably associated with the amorphous phase of the polymer and affected by the cooperative motion of polymer segments and by the free volume. Finally, the glass transition of $\alpha$-PVDF can be independently determined from the conductivity with the actual and modified VFT expressions of Eqs.~(\ref{eq:3}) and (\ref{eq:5}), respectively.



We would like to thank Dr. Wolgang K{\"u}nstler for his valuable comments and providing the raw material.



\end{document}